\documentclass{PoS}
\usepackage[utf8]{inputenc}
\usepackage{amsmath,mathtools} 

\newcommand{\R}{\mathbb{R}}
\newcommand{\cP}{\mathcal{P}}
\newcommand{\cB}{\mathcal{B}}
\newcommand{\cH}{\mathcal{H}}
\newcommand{\np}{\emptyset}
\newcommand{\zero}{\mathbf{0}}
\newcommand{\pcomp}{\diamond}

\title{A f{}irst-principles approach to physics based on locality and operationalism}

\ShortTitle{A first-principles approach to physics based on locality and operationalism}

\author{\speaker{Robert Oeckl}%
         \thanks{This work was supported in part by UNAM-DGAPA-PASPA through a sabbatical grant, by UNAM-DGAPA-PAPIIT through project grant IN100212 and by the Emerging Fields project ``Quantum Geometry'' of the Friedrich-Alexander-Universität Erlangen-Nürnberg.}\\
        Centro de Ciencias Matemáticas, Universidad Nacional Autónoma de México, Morelia, Mexico\\
        E-mail: \email{robert@matmor.unam.mx}}

\abstract{Starting from the guiding principles of spacetime locality and operationalism, a general framework for a probabilistic description of nature is proposed. Crucially, no notion of time or metric is assumed, neither any specific physical model. Remarkably, the emerging framework converges with a recently proposed formulation of quantum theory, obtained constructively from known quantum physics. At the same time the framework also admits statistical theories of classical physics.}

\FullConference{Frontiers of Fundamental Physics 14\\
		 15-18 July 2014\\
		 Aix Marseille University (AMU) Saint-Charles Campus, Marseille, France}

\begin{document}

\section{Introduction}

In this short note we shall sketch a particular first-principles approach to the structure of a physical description of nature. That is, we shall outline the development of a system of mathematical objects together with specifications of how these objects relate to nature in principle. This is to be done in such a way that the emerging framework may accommodate the same predictive power we are familiar with from known physical theories.
However, we shall presuppose neither any specific physical theory nor even merely the mathematical structure of a given physical theory or framework (such as, in particular, quantum theory).
Of course, there cannot be hope for success if we keep this endeavor entirely divorced from physical experience.
Indeed, we shall rely on experience condensed from the most comprehensive physical descriptions of nature known. We extract two key principles that turn out to have far reaching consequences when combined with generic probabilistic reasoning and will be sufficient for our purposes.\footnote{We are far from claiming there are no further principles to be derived from these theories, or even that the ones we choose are to be considered the most important ones.} These principles are \emph{locality} and \emph{operationalism}.

The principle of locality stems from the 19th century discovery that forces do not act at a distance, but are mediated through fields that permeate spacetime. In classical physics particles can only interact through signals carried by fields that connect them in spacetime. This remains true in essence in quantum theory, although particles and fields are replaced there (in quantum field theory) by a unified notion. Thus, interaction can only happen through adjacency in spacetime and is parametrizable through possible ``signals''.

The principle of operationalism is motivated by the discovery of quantum physics. In classical physics reality is described through particle trajectories and field configurations distributed in spacetime. This distribution is an objective fact that exhaustively describes reality and does not depend on the observer or its actions. But we have learned that this is not an accurate description of nature and invented quantum theory to provide a better description. In quantum theory the process of measurement and the observer play a distinguished role. The lesson is that rather than trying to describe an abstract reality that we are somehow in contact with, we should concentrate on the contact itself. That is, we should describe reality \emph{through} the act of probing it, i.e., through measurement, observation, preparation etc. This is what we mean by operationalism.

\section{Spacetime}

A manifestation of locality essential for the successful physical description of our world is the fact that an experiment performed in a laboratory generally does not depend (in important ways and if it is not so intended) on what happens outside the laboratory and vice versa. To implement this aspect of locality in our framework we need a manner to distinguish the laboratory from the rest of the universe. That is accomplished by a notion of spacetime \emph{region}. It serves precisely as a means to distinguish whatever happens inside the region from the rest of the universe.
A crucial element of physical theory consists in establishing relations between what happens in one region and another one. By locality, the interior of a region can interact with the rest of the universe only through the region's \emph{boundary}. Thus, a notion of \emph{adjacency} of regions is required. More technically speaking this means a notion of gluing or \emph{composing} regions along common boundary parts. These boundary parts are spacetime \emph{hypersurfaces}.
We require thus two types of primitive spacetime objects: regions and hypersurfaces, with boundaries of regions being a special case of hypersurfaces. To be mathematically specific we take the regions to be 4-dimensional topological manifolds and the hypersurfaces to be 3-dimensional topological manifolds. In addition to the spacetime objects themselves there is a notion of \emph{composition} of regions along hypersurfaces. Mathematically, this is a gluing of manifolds. Also we require a notion of \emph{decomposition} of hypersurfaces into component hypersurfaces since the gluing is, in general, only along parts of boundaries. For our present purposes it is not necessary, however, to make this mathematically more precise.

\section{Probes}
\label{sec:probes}

By operationalism, rather than aiming for some ontological description of physics, we should describe the act of experimentation, observation, preparation etc.\ together with its outcomes (if any). We subsume this in the concept of a \emph{probe}. In order to make use of locality, a probe is assigned to a spacetime region. It encodes direct influence on, and outcomes of observations in, only this spacetime region. A probe may correlate to events in other spacetime regions, but, due to locality, exclusively so through ``signals'' traversing the boundary of the region. Mathematically, to any spacetime region $M$ we assign a set of probes $\cP_M$ in $M$. For any region $M$, there is a special \emph{null-probe} $\np\in\cP_M$. This encodes leaving the region ``empty'', i.e., not making any observation in $M$, not putting any apparatus there etc. An elementary, but important operation on probes is their \emph{composition}. Say there are regions $M$ and $N$ that can be composed to a joint region $M\cup N$. Then a probe in $N$ together with a probe in $M$ determine a probe in $M\cup N$. That is, there is a composition map $\pcomp:\cP_M\times\cP_N\to\cP_{M\cup N}$. Physically speaking this is the triviality that the combination of two experiments is also an experiment.

\section{Boundary conditions}

As stated, we take locality to imply that whatever happens inside a region influences and is influenced by the rest of the universe exclusively through ``signals'' that traverse the region's boundary. We condense this into a notion of \emph{boundary conditions}. That is, whatever happens outside of the region can be parametrized in so far as it affects the interior by specifying a boundary condition. The same holds for the influence of the interior on the exterior. Moreover, again using locality, boundary conditions should be localizable not only on boundaries, but also on pieces of them, i.e., on general hypersurfaces. Mathematically, we associate to any hypersurface $\Sigma$ a set of boundary conditions $\cB_{\Sigma}$.

\section{Values}

So far, we have limited ourselves to a qualitative description of the ingredients of our framework. To make actual predictions we need numbers. We shall follow tradition in physics and assume that any observation or measurement outcome can be described through a finite number of real numbers. This motivates the notion of \emph{values}, refining at the same time the notions of probes and boundary conditions. To a spacetime region $M$, a probe $P\in\cP_M$ and a boundary condition $b\in\cB_{\partial M}$ we assign a value, i.e., a real number. We shall use the notation $(P,b)_M$ to denote this value. Thus, the value represents the outcome for probe $P$ in region $M$ given the boundary condition $b$. Since a value is a single real number, an experiment might not be representable by a single probe, but might require a number of probes for its description.
Crucially, the assignment of values means that for a given region $M$ we may view a probe $P\in\cP_M$ as a real valued function $(P,\cdot)_M$ on the set $\cB_{\partial M}$ of boundary conditions. Similarly, we may view a boundary condition $b\in\cB_{\partial M}$ as a real valued function $(\cdot,b)_M$ on the set $\cP_M$ of probes. In particular, both the set of probes and the set of boundary conditions naturally span real vector spaces of functions. However, the sets are in general not identical to the vector spaces. Arbitrary linear combinations of boundary conditions are not necessarily again boundary conditions. The same goes for probes.

\section{Hierarchies of probes and partial order}

Another important structure that the sets of probes acquires by virtue of being a set of functions is a \emph{partial order}. Given two probes $P,Q\in\cP_M$ in a region $M$ we declare $P\le Q$ if and only if for any boundary condition $b\in\cB_{\partial M}$ we have, $(P,b)_M\le (Q,b)_M$. This partial order carries important physical meaning as we illustrate in the following.
To simplify the argument we imagine for a moment that a value predicts directly the reading on an apparatus in an experiment. (We shall see later that this cannot be the case in general.) Consider probes that represent experiments that have YES/NO outcomes. That is, these probes yield values in the set $\{0,1\}$, with $0$ representing NO and $1$ YES. We shall call these \emph{primitive probes}. The null-probe is then a primitive probe that always yields YES.
Imagine an apparatus that displays a single light in such a way that the light might either show red or green, nothing else. There are different primitive probes associated with the apparatus. There is the probe that yields $1$ if the light shows green, and $0$ otherwise. Call this $P(g)$. Similarly, there is the probe that yields $1$ if the light shows red, and $0$ otherwise. Call this $P(r)$. There is also the probe that encodes the mere presence of the apparatus, without considering what color the light shows, call this $P(*)$. The probe $P(*)$ is more general than the probes $P(g)$ and $P(r)$ as the latter correspond to special configurations of the former. This \emph{hierarchy} of generality is reflected in a \emph{partial order} relation on those probes. For any boundary condition, the induced values for the probes satisfy the same order relations which are thus order relations between the probes, namely,
\begin{equation}
 \zero\le P(g) \le P(*)\quad\text{and}\quad \zero\le P(r) \le P(*) .
\label{eq:ex1po}
\end{equation}
Here $\zero$ represents the trivial probe that always returns the value $0$. Apart from the order relation we also have an additivity relation here, namely $P(*)=P(r)+P(s)$.
In this simple example the hierarchy and partial order are rather limited. But with growing complexity of the apparatus, the hierarchies quickly become more interesting and powerful. As a second example take an apparatus that displays two lights, each either showing red or green. With the obvious notation we have, for example,
\begin{equation}
 \zero\le P(g,r) \le P(*,r)\le P(*,*),
\end{equation}
and a number of similar relations. The partially ordered set of the associated primitive probes is considerably richer than in the first example. Partial orders also arise from continuous measurement outcomes. Consider an apparatus with a scale, showing numbers from $0$ to $5$. The primitive probe that represents an outcome in the range $[0,4]$ is more general than the primitive probe that represents an outcome in the range $[1.5, 3.5]$ for example, etc.

\section{Probability and Conditionality}

The assumption that values always directly predict measurement outcomes has very restrictive implications. Firstly, it implies that any boundary condition is compatible with any apparatus and also with the absence of an apparatus (encoded by the null-probe). This implication is not at all innocent. Indeed, by locality the set of boundary conditions associated to a hypersurface does not ``know'' about the restrictions imposed by the presence of a region much less an apparatus in a region that the hypersurface might be the boundary of. This suggests that the assumption, at least in this simple form, is unsustainable. A second implication is that measurement outcomes are always predicted with certainty. Our experience with quantum physics speaks against this. A third implication is that boundary conditions need to be mutually exclusive, which is highly restrictive.

We should thus allow values to have a probabilistic and conditionalistic relation to measurement outcomes. This requires to adapt the concept of primitive probes. While we continue to consider probes that encode YES/NO measurement outcomes as primitive, we cannot restrict them to yield the binary values $0$ or $1$. Even restricting them to values in the interval $[0,1]$ could meet with normalizability issues. So we merely require them to yield non-negative values. As we shall see, this lack of normalization is not a problem. We also allow probes to be formed as probabilistic ensembles of other probes. These are linear combinations of probes with positive coefficients summing to $1$. While they might not represent a given single experiment, they can give rise to meaningful statements about ensembles. Combining this with the fact that numbers associated to measurement outcomes can be redefined simply by convention justifies considering arbitrary linear combinations of probes as (general, not necessarily primitive) probes. We therefore take the set of probes $\cP_M$ from here onward to form a real vector space. The primitive probes admit linear combinations with positive coefficients by forming ensembles and relaxing normalization. Thus, they form a \emph{positive cone} that we shall denote $\cP^+_M$ in $\cP_M$, making $\cP_M$ into an \emph{ordered vector space}.

Crucially, the discussion of the previous section on the physical meaning of hierarchies of probes remains valid in the probabilistic context. In particular, the partial order relations in the examples remain true. However, rather than directly yielding a value corresponding to YES/NO the probes in question yield (relative) probabilities. Recall the example with the apparatus displaying a single light with states ``red'' or ``green''. Given a boundary condition $b\in\cB_{\partial M}$ the value of $(P(g),b)_M$ is not restricted to the set $\{0,1\}$, corresponding to ``not green'' or ``green''. But neither is it in general the \emph{probability} for the state to turn out ``green'' in the experiment. Rather, this probability is given by the \emph{quotient}
\begin{equation}
\frac{(P(g),b)_M}{(P(*),b)_M} .
\label{eq:condpp}
\end{equation}
That is, to obtain the probability we have to \emph{condition on} the presence of the apparatus.
The inequality (\ref{eq:ex1po}) precisely guarantees that this quotient lies in the interval $[0,1]$.
The value $(P(*),b)_M$ itself can be seen as a measure of the ``compatibility'' between the boundary condition $b$ and the presence of the region $M$ with the apparatus represented by the probe $P(*)$. Similarly, the value $(\np,b)_M$ associated to the null-probe measures the ``compatibility'' between the boundary condition $b$ and the presence of the region $M$.
Now recall, from last section, the apparatus displaying two lights, each showing either red or green. In this case we can express non-trivial probabilities relating two apparatus states. For example, with boundary condition $b\in\cB_{\partial M}$ the probability that the first light shows green given that the second light shows green is given by the quotient,
\begin{equation}
\frac{(P(g,g),b)_M}{(P(*,g),b)_M} .
\end{equation}

\section{Hierarchies of boundary conditions}

In the probabilistic setting boundary conditions need not be mutually exclusive, may be probabilistically combined, and become subject to the formation of hierarchies. As a physical example consider temperature (range) as a boundary condition. For example, a temperature range $b$ between $10$ and $20$ degrees Celsius is more general than one $c$ between $10$ and $15$ degrees. The corresponding partial order is $c\le b$. However, this does \emph{not} reflect an order relation between values $(P,c)_M\le (P,b)_M$ for arbitrary probes $P\in\cP_M$. It does reflect this order relation between values, however, for the restricted class of primitive probes $P\in\cP_M^+$. As in the case of probes we do not assume normalizability. Thus, any linear combination of boundary conditions with positive coefficients is again a valid boundary condition. We modify our notation slightly and denote the set of boundary conditions by $\cB^+_{\Sigma}$ while $\cB_{\Sigma}$ shall henceforth denote the real vector space spanned by it. Thus, $\cB^+_{\Sigma}$ is a positive cone in $\cB_{\Sigma}$, which thus acquires the structure of an ordered vector space.

Analogous to conditional probabilities for different probes given a boundary condition, it makes equal sense to consider conditional probabilities relating different boundary conditions for a given probe. For boundary conditions $b,c\in\cB^+_{\partial M}$ with $\zero\le c\le b$, i.e., $c$ a specialization of $b$, we can ask for the probability that $c$ is realized given that we know $b$ to be realized and given the primitive probe $P\in\cP^+_M$ in $M$. This probability is the quotient,
\begin{equation}
 \frac{(P,c)_M}{(P,b)_M} .
\label{eq:condbdy}
\end{equation}
The physical information added that gives meaning to this probability is the presence of the region $M$ and associated restriction as to what may happen at its boundary $\partial M$.

\section{Expectation values}

For an expectation value two probes are required: One probe $Q_0$ that encodes the mere presence of the measurement apparatus without considering the outcome of the measurement and another probe $Q$ that takes account of the outcome of the measurement. The expectation value is thus,
\begin{equation}
\frac{(Q,b)_M}{(Q_0,b)_M} ,
\label{eq:condev}
\end{equation}
in complete analogy to formula (\ref{eq:condpp}) for primitive probes.

\section{Composition}
\label{sec:composition}

We return to the concept of composition of probes introduced in Section~\ref{sec:probes}. It turns out that locality, suitably understood, is sufficient to derive a \emph{composition law} for probes. We give a condensed account of this in the following. As a first step we introduce \emph{slice regions}. These are hypersurfaces $\Sigma$ considered as infinitesimally thin regions $\hat{\Sigma}$. A slice region has a boundary $\partial\hat{\Sigma}$ with two components $\Sigma$, $\Sigma'$, each of which is a copy of the original hypersurface. The null-probe gives rise to a bilinear map $\cB_{\Sigma}\times\cB_{\Sigma}\to\R$ via $(b_1,b_2)\mapsto (\np,(b_1,b_2))_{\hat{\Sigma}}$. Assuming that different boundary conditions encode different physics, this inner product must be \emph{non-degenerate}. It may have \emph{positive definite} and \emph{negative definite} parts.\footnote{We deliberately ignore here complications that may arise in an infinite-dimensional setting.} An orthonormal basis $\{b_k\}_{k\in I}$ has the property, $(\np, (b_k , b_l))_{\hat{\Sigma}} = (-1)^{\sigma(k)} \delta_{k,l}$. Here $\sigma(k)=0,1$ for the positive-definite and negative-definite part respectively.
Now, consider the composition of probes $P$, $Q$ in adjacent spacetime regions $M$, $N$. We fix moreover boundary conditions $b$ on $\partial P\setminus \partial Q$ and $c$ on $\partial Q\setminus \partial P$. By locality, it must be possible to describe the effect of probe $Q$ in $N$ with $c$ on $P$ equivalently through a boundary condition $q$ on the interfacing hypersurface $\Sigma = \partial M \cap \partial N$. Formally, $(P\pcomp Q, (b, c))_{M\cup N} = (P, (b, q))_M$. Specializing to the case that $M$ is a slice region $\hat{\Sigma}$ with $P$ the null-probe we get,
\begin{equation}
(\np, (b, q))_{\hat{\Sigma}} =
(Q, b)_N = \sum_k (-1)^{\sigma(k)} (\np, (b, b_k ))_{\hat{\Sigma}} (Q, b_k)_N
\end{equation}
For the second equality have used a completeness relation for the inner product. But this implies, $q=\sum_k (-1)^{\sigma(k)} b_k (Q, b_k)_N$. This in turn implies the composition rule,
\begin{equation}
(P\pcomp Q, (b, c))_{M\cup N}
= \sum_k (-1)^{\sigma(k)} (P, (b, b_k ))_{M} (Q, (b_k , c))_{N} .
\label{eq:comprule}
\end{equation}

\section{Convergence}

Remarkably, the framework we have arrived at turns out to be essentially a formulation of quantum theory. To explain this, we recall the \emph{general boundary formulation} of quantum theory \cite{Oe:gbqft}, which incorporates manifest locality into quantum theory. In its \emph{amplitude formalism (AF)}, (generalized) Hilbert spaces of states are associated to spacetime hypersurfaces. Amplitudes and observables are associated to spacetime regions. The AF incorporates a composition rule similar to (\ref{eq:comprule}), coming from \emph{topological quantum field theory}. The \emph{positive formalism (PF)} \cite{Oe:dmf} is obtained from the AF by generalizing from pure states to mixed states. It turns out that the framework we have arrived at in this note coincides with the PF. More precisely, the ordered vector space $\cB_{\Sigma}$ of boundary conditions arises in the PF as the space of self-adjoint operators on the Hilbert space $\cH_{\Sigma}$ (of the AF) associated with the hypersurface $\Sigma$. The primitive probes arise as (unnormalized) \emph{quantum operations}. General probes are quantum observables and quantum measurements. The composition rule (\ref{eq:comprule}) is exactly a version of the Axioms (P5a), (P5b) in \cite{Oe:dmf}. Transition probabilities arise as special cases of formula (\ref{eq:condbdy}) and expectation values as special cases of formula (\ref{eq:condev}). If, as usual in the standard formulation of quantum theory, spacetime regions are time intervals and temporally final states are conditioned on initial ones, the denominators in these formulas turn out to be equal to $1$. This explains why in the standard formulation, transition probabilities and expectation values do not normally take the form of quotients.

In \cite{Oe:dmf} it was shown that even though spaces of self-adjoint operators (for the boundary conditions) have more structure than that of ordered vector spaces, only the latter is necessary for a coherent and predictive physical framework. Indeed, in this note we have arrived only at ordered vector spaces. It turns out that dropping the more restrictive structure of self-adjoint operators is akin to dropping the restriction to quantum theory, as we shall see in a moment. At this point we can only speculate what exactly makes the difference between classical and quantum theory. It might be the difference between boundary conditions forming a lattice versus an anti-lattice.

\section{Bonus: Classical physics}

We proceed to show how classical physics can also be accommodated fairly naturally within the presented formalism. A classical theory associates a space of solutions $L_M$ of the equations of motion to each region $M$ and of germs of solutions $L_{\Sigma}$ to each hypersurface $\Sigma$. Note that for any region $M$ we have a map $L_M\to L_{\partial M}$ that consists in retaining the solution only on the boundary. We can now set up a \emph{deterministic} version of the framework as follows: We define the space of boundary conditions to be $\cB_{\Sigma}=L_{\Sigma}$. (We return momentarily to the setting where $\cB_{\Sigma}$ is a set without necessarily having the structure of a vector space.) We define the null-probe $(\np,b)_M$ to take the value $1$ if there is a solution in $M$ that induces the boundary condition $b$ and $0$ otherwise. We take general probes to be induced by local observables. That is given an observable $O:L_M\to\R$ in a spacetime region we define an associated probe $P_O$ via $(P_O,b)_M=O(\phi)$ if there is a solution $\phi\in L_M$ that reduces to $b\in L_{\partial M}$ on the boundary and as $0$ otherwise. (This supposes that different solutions in the interior can be distinguished on the boundary or that we restrict to observables that give the same value if the induced boundary conditions are identical.) Since the spaces of boundary conditions are not in general vector spaces here, the composition rule of Section~\ref{sec:composition} does not apply.

A probabilistic setting is achieved by transitioning to classical statistical physics. Thus, the space of boundary conditions is replaced by the space of probability distributions over $L_{\Sigma}$. Not imposing normalization, this becomes a cone $\cB^+_{\Sigma}$ of positive distributions inside a real vector space $\cB_{\Sigma}$. Integrating over $L_M$ in the interior with such a distribution makes a probe as defined above for the deterministic setting into one for this statistical setting. (We omit details here which may render converting this sketch into a precise formalism quite non-trivial.) In this case the composition rule of Section~\ref{sec:composition} should apply. The present framework, along the lines sketched, might provide a suitable starting point for a statistical treatment of classical field theory without metric background, such as general relativity.

\bibliographystyle{pos} 
\bibliography{stdrefsb}

\begin{thebibliography}{1}
\providecommand{\url}[1]{\texttt{#1}}
\providecommand{\urlprefix}{URL }
\providecommand{\selectlanguage}[1]{\relax}
\providecommand{\eprint}[2][]{[#2]}

\bibitem{Oe:gbqft}
R.~Oeckl, \textit{General boundary quantum field theory: Foundations and
  probability interpretation}, Adv. Theor. Math. Phys. \textbf{12} (2008)
  319--352 \eprint{hep-th/0509122}

\bibitem{Oe:dmf}
R.~Oeckl, \textit{A Positive Formalism for Quantum Theory in the General
  Boundary Formulation}, Found. Phys. \textbf{43} (2013) 1206--1232
  \eprint{1212.5571}

\end{thebibliography}

\end{document}